\documentclass[11pt,twoside]{article}

\usepackage[numbers,square]{natbib}
\usepackage[utf8]{inputenc}
\usepackage{graphicx}
\usepackage{amsmath}
\usepackage{amssymb}
\usepackage{url}
\usepackage{graphics}
\usepackage{wasysym}
\usepackage{enumitem}
\usepackage[english]{babel}
\usepackage{color}
\usepackage[usenames,dvipsnames]{xcolor}

\begin{document}

%*********************************************************************** TITULO

\section*{Asteroseismology of Exoplanet-Host Stars in the \emph{Kepler} Era}

\vspace{10pt}

%************************************************************************* AUTORES
Tiago L.~Campante$^{1}$

\vspace{10pt}

%************************************************************************** MORADA

{
\center

    \footnotesize

$^{1}$ School of Physics and Astronomy, University of Birmingham, Edgbaston, Birmingham B15 2TT, United Kingdom \\
E-mail: campante@bison.ph.bham.ac.uk

}

\vspace{25pt}

%%%%%%%%%%%%%%%%%%%%%%%%%%%%%%%%%%%%%%%%%% RESUMO

{\setlength{\parindent}{30pt}%

\small%
\section*{Abstract}
\smallskip
%%%%%%%%%%%%%%%%
New insights on stellar evolution and stellar interior physics are being made possible by asteroseismology, the study of stars by the observation of their natural, resonant oscillations. Asteroseismology is making significant contributions to our understanding of solar-type stars, in great part due to the exquisite data that have been made available by NASA's \emph{Kepler} space telescope. Of particular interest is the synergy between asteroseismology and exoplanetary science. Herein I will review recent contributions from asteroseismology to the determination of fundamental properties of \emph{Kepler} exoplanet-host stars and stress its potential in constraining the spin-orbit alignment of exoplanet systems.

\vspace{15pt}

%%%%%%%%%%%%%%%%%%%%%%%%%%%%%%%%%%%%%%%%%%%    SECTIONS  %%%%%
\section{Introduction}
Solar-type stars show rich spectra of oscillations, which are excited and intrinsically damped by turbulence in the outermost layers of their convective envelopes. Prior to the advent of \emph{Kepler} \citep{Kepler}, solar-like oscillations had been detected in only a few tens of main-sequence and subgiant stars using ground-based high-precision spectroscopy or wide-field photometry from the \emph{CoRoT} space telescope \citep{CoRoT}. Photometry from \emph{Kepler} has since led to an order of magnitude increase in the number of such stars with confirmed oscillations \citep{KeplerEnsemble}.

The information contained in solar-like oscillations allows fundamental stellar properties (i.e., density, surface gravity, mass, radius, and age) to be precisely determined. Furthermore, the internal stellar structure can be constrained to unprecedented levels, provided that individual oscillation mode parameters are measured. As a result, asteroseismology of solar-type stars is quickly maturing into a powerful tool and its impact is being felt more widely across different domains of astrophysics. A noticeable example is the synergy between asteroseismology and exoplanetary science. Planetary-transit observations -- as carried out by \emph{Kepler} -- are an indirect detection method, and are consequently only capable of providing planetary properties relative to the properties of the host star. The precise characterization of the host star through asteroseismology thus allows for inferences on the properties of its planetary companions. Information on the stellar internal rotation and inclination angle as provided by asteroseismology can also lead to a better understanding of the planetary system dynamics and evolution. Moreover, the potential use of asteroseismology in measuring the levels of near-surface magnetic activity and in probing stellar activity cycles may help constrain the location of habitable zones around solar-type stars.

\section{Precise characterization of exoplanet-host stars}
The first application of asteroseismology to known exoplanet hosts in the \emph{Kepler} field \citep{JCD10} was followed by a series of planet discoveries where asteroseismology was used to constrain the properties of the host star. A systematic study of \emph{Kepler} planet-candidate host stars using asteroseismology was presented in Huber et al.~\citep{HuberExo}. Fundamental properties were determined for 66 host stars based on average asteroseismic parameters, with typical uncertainties of $3\,\%$ and $7\,\%$ in radius and mass, respectively. Combined with an estimate of $R_{\rm p}/R_{\rm s}$ (i.e., the planet-to-star radius ratio) from transit photometry, this yielded precise radii ($\sim\!3\,\%$) for over 100 exoplanet candidates, while raising the number of \emph{Kepler} hosts with asteroseismic solutions to nearly 80 stars. 

A detailed modeling of the frequencies of oscillation can lead to considerably more precise stellar properties. Kepler-10, \emph{Kepler}'s first rocky-exoplanet host, had its radius measured with a precision better than $1\,\%$ from detailed frequency modeling \citep{Kepler-10}, with the radius of Kepler-10b given to within just $125\:{\rm km}$. Even in the event of not being able to measure individual frequencies, a mere detection of the periodicity of the pattern of solar-like oscillations is sufficient to estimate the stellar radius. Perhaps the most notorious example has been the characterization of the solar analog Kepler-22 \citep{Kepler-22}, host to the first known transiting planet to orbit within the habitable zone of a Sun-like star.

When combined with transit photometry and follow-up radial-velocity observations, the asteroseismic mass yields the absolute planetary mass. In the absence of radial-velocity data, transit-timing variations (TTVs), caused by mutual gravitational interactions in multiple-planet systems, may be used to constrain the planetary mass. The first instance of the use of asteroseismology in combination with TTVs by means of a photodynamical model occurred in the analysis of the Kepler-36 system \citep{Kepler-36}. The planetary radii and masses of the two transiting planets in the system were constrained to better than $3\,\%$ and $8\,\%$, respectively. The planets, a super-Earth and a mini-Neptune, are unusual in that their orbital distances differ by only $10\,\%$ while their densities differ by a factor of eight.

The asteroseismic age provides an upper-limit age estimate for the planets and can be used to assess the dynamical stability of the system. The parent star in the Kepler-444 system has an age of $11.2\pm1.0\:{\rm Gyr}$ from detailed frequency modeling, making it the oldest known system of terrestrial-size planets \citep{Kepler-444}. The precision with which the age of Kepler-444 has been determined from asteroseismology ($\sim\!9\,\%$) is an impressive technical achievement that was only made possible due to the
extended and high-quality photometry provided by the \emph{Kepler} mission. We have thus attained the level of precision expected for ESA's \emph{PLATO} mission, which has the science goal of providing stellar ages to $10\,\%$ precision as a key to exoplanet parameter accuracy \citep{PLATO}.

\section{Spin-orbit alignment of exoplanet systems}
The angle $\psi$ between the planetary orbital axis and the stellar spin axis is a fundamental geometric property, having been recognized as an important diagnostic of theories of planet formation, migration, and tidal evolution. However, $\psi$ is not directly measurable, making it a matter of the utmost importance to seek for empirical constraints on $\psi$. Asteroseismology provides a potentially powerful means of directly determining the angle $i_{\rm s}$ between the stellar spin axis and the line of sight. The asteroseismic estimation of $i_{\rm s}$ rests on our ability to resolve and extract signatures of rotation in the frequency-power spectra of non-radial modes of oscillation. The applicability of this technique depends entirely on the stellar properties and not on the planetary or orbital parameters, which can be regarded as a clear advantage when measuring obliquities of systems with small and/or long-period planets. This technique has recently been applied to several Sun-like exoplanet hosts observed with \emph{Kepler} \citep{ChaplinObl,Lund}.

%%%%%%%%%%%%%%%%%%%%%%%%%%% FIGURA
\begin{figure}[!t]
\centering
\includegraphics[scale=0.4]{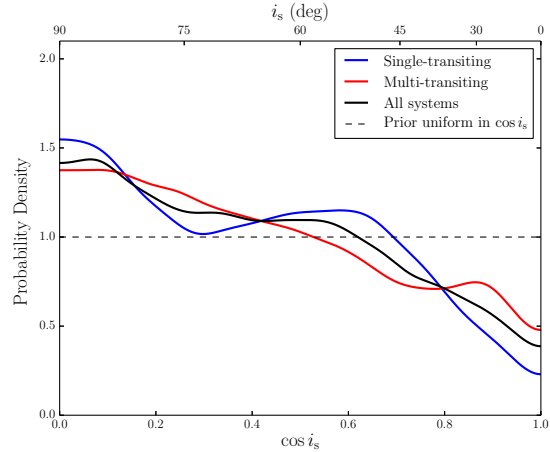}
\caption{\footnotesize Average posterior probability distribution of $\cos i_{\rm s}$ (kernel density estimate) for stars in the asteroseismic sample. For reference, the dashed horizontal line represents the (uninformative) isotropic prior on $i_{\rm s}$ (or, equivalently, the uniform prior on $\cos i_{\rm s}$) adopted in the asteroseismic analysis.\label{fig:avrpost}}
\end{figure}

For a transiting system, a small value of $i_{\rm s}$ implies a spin-orbit misalignment. The converse is not true, since a large value of $i_{\rm s}$ is consistent with but does not necessarily imply a spin-orbit alignment. The interpretation of spin-orbit alignment in terms of the measured $i_{\rm s}$ can thus be ambiguous and we require a statistical analysis of an ensemble of such measurements to draw general inferences \citep{MortonWinn}. The first statistical analysis is currently underway of an ensemble of asteroseismic obliquity measurements obtained for Sun-like stars hosting transiting planets (Fig.~\ref{fig:avrpost}; T.~L.~Campante et al., in preparation). 

Large obliquities are frequently observed for systems harboring hot Jupiters, whereas compact multi-transiting systems seem to preferentially display low obliquities \citep{Albrecht13}. Consequently, migration mechanisms capable of exciting large obliquities now start being favored over the paradigm of inward spiraling along the protoplanetary disk to explain hot-Jupiter formation. This assumes that the protoplanetary disk is coplanar with the stellar equator. However, the possibility remains that primordial star-disk misalignments are ubiquitous, meaning that large obliquities could be a generic feature of planetary systems and not specific to hot-Jupiter formation. This hypothesis may in principle be tested by measuring the obliquities of systems with multiple transiting planets, since the planetary orbits in these systems are nearly coplanar and presumably trace the plane of the protoplanetary disk. The ongoing statistical analysis of an ensemble of asteroseismic obliquity measurements is expected to provide further clues on this matter.

%%%%%%%%%%%%%%%%%%%%%%%%%%%%%%%%%%%
%%
%%                 AKNOWLEDGMENTS & BIBLIOGRAPHY
%%
%%%%%%%%%%%%%%%%%%%%%%%%%%%%%%%%%%%

{\scriptsize%

%%%%%%%%%%%%%%%%%%%%%%%%%%%%%%%%%%%%%%%%%%%    AGRADECIMENTOS  %%%%%
%\section*{Acknowledgments}
%Text

%%%%%%%%%%%%%%%%%%%%%%%%%%%%%%%%%%%%%%%%%%%    REFERÊNCIAS  %%%%%
%\section*{References}
%\vspace{5pt}

\bibliographystyle{naturemag}
\bibliography{biblio}

%\begin{enumerate}[label={[\arabic*]}]

%\item

%\item

%\item

%\end{enumerate}

}

\end{document}